\documentclass[prb,preprint,showpacs,amsmath,amssymb,superscriptaddress]{revtex4}
\usepackage{epsfig}
\usepackage{graphicx}
\usepackage{amsmath,amssymb}

\newcommand{\beq}[1]{
\begin{equation}
\label{e#1} }

\newcommand{\eeq}{
\end{equation}
}

\begin{document}

\title{Tunneling anisotropic magnetoresistance of NiFe/IrMn/MgO/Pt stack: An antiferromagnet based spin-valve}

\author{B.~G.~Park}
\affiliation{Hitachi Cambridge Laboratory, Cambridge CB3 0HE, United Kingdom}

\author{J.~Wunderlich}
\affiliation{Hitachi Cambridge Laboratory, Cambridge CB3 0HE, United Kingdom}
\affiliation{Institute of Physics ASCR, v.v.i., Cukrovarnick\'a 10, 162 53
Praha 6, Czech Republic}

\author{X.~Mart\'{i}}
\author{V.~Hol\'y}
\affiliation{Faculty of Mathematics and Physics, Charles University in Prague, Ke Karlovu 3, 121 16 Prague 2, Czech Republic}

\author{Y.~Kurosaki}
\author{M.~Yamada}
\author{H.~Yamamoto}
\author{A.~Nishide}
\author{J.~Hayakawa}
\affiliation{Hitachi Ltd, Advanced Research Laboratory, 1-280 Higashi-koigakubo, Kokubunju-shi, Tokyo 185-8601, Japan}
\author{H. Takahashi}
\affiliation{Hitachi Cambridge Laboratory, Cambridge CB3 0HE, United Kingdom}
\affiliation{Hitachi Ltd, Advanced Research Laboratory, 1-280 Higashi-koigakubo, Kokubunju-shi, Tokyo 185-8601, Japan}
\author{A.~B.~Shick}
\affiliation{Institute of Physics ASCR, v.v.i., Na Slovance 2, 182 21 Praha 8, Czech Republic}

\author{T.~Jungwirth}
\affiliation{Institute of Physics ASCR, v.v.i., Cukrovarnick\'a 10, 162 53
Praha 6, Czech Republic} \affiliation{School of Physics and
Astronomy, University of Nottingham, Nottingham NG7 2RD, United Kingdom}
\date{\today}

\maketitle

{\bf 
Spin-valve is a microelectronic device in which high and low resistance states are realized by utilizing both charge and spin of carriers. Spin-valve structures used in modern hard drive read-heads and magnetic random access memories comprise two ferromagnetic (FM) electrodes whose relative magnetization orientations can be switched between parallel and antiparallel configurations, yielding the desired giant or tunneling magnetoresistance effect.\cite{Chappert:2007_a} In this paper we demonstrate $>100$\% spin-valve-like signal in a NiFe/IrMn/MgO/Pt stack with an antiferromagnet (AFM) on one side and a non-magnetic metal on the other side of the tunnel barrier. FM moments in NiFe are reversed by external fields $\lesssim50$~mT and the exchange-spring effect\cite{Scholl:2004_a} of NiFe on IrMn induces rotation of AFM moments in IrMn which is detected by the measured tunneling anisotropic magnetoresistance (TAMR).\cite{Shick:2010_a} Our work demonstrates a spintronic element whose transport characteristics are governed by an AFM. It demonstrates that sensitivity to low magnetic fields can be combined  with large, spin-orbit coupling induced magneto-transport anisotropy using a single magnetic electrode. The AFM-TAMR provides means to study magnetic characteristics of AFM films by an electronic transport measurement. 
}

AFMs represent the overwhelming majority of magnetically ordered materials. Yet, in current microelectronics  they only play a static supporting role of enhancing the magnetic hardness of FM electrodes via the exchange-bias effect.\cite{Nogues:1999_a,Chappert:2007_a} Studies of magnetoresistive phenomena potentially suitable for spintronics, in which transport is governed by AFMs, have focused on  counterparts of conventional  giant or tunneling magnetoresistance devices with two magnetic electrodes.\cite{Nunez:2006_a,Haney:2006_b,Duine:2007_a} It has been acknowledged that  requirements on the structural quality and the coherence of transport through interfaces in these types of devices are significantly more stringent in the case of AFMs and experimental realization of this concept remains a great challenge. 

Recent {\em ab initio} theory studies,\cite{Shick:2010_a,Jungwirth:2010_a} co-authored by several of us, have predicted an alternative route towards AFM spintronics based on relativistic spin-orbit coupling phenomena, in particular on the TAMR. The effect relies on the variation of the orientation of moments in a single magnetic electrode, circumventing the stringent requirements on the coherence of spin-dependent transport in structures with two or multiple magnetic electrodes. To date, TAMR has been experimentally demonstrated only in FMs.\cite{Gould:2004_a,Brey:2004_b,Shick:2006_a,Gao:2007_a,Park:2008_a} However, as proposed in Ref.~\onlinecite{Shick:2010_a}, it should be equally well present in AFMs. Moreover, the magnetic anisotropy phenomena are maximized in bimetallic systems combining large spontaneous moments on the 3$d$ shell of a transition metal and large magnetic susceptibility and spin-orbit coupling on the 5$d$-shell of a noble metal.\cite{Wunderlich:2006_a,Shick:2008_a,Park:2008_a} Since Mn carries the largest  moment among transition metals and most of the bimetallic alloys containing Mn order antiferromagnetically, the goals of strong magnetic anisotropy phenomena and of AFM spintronics  appear to merge naturally together. These predictions are substantiated in our experiments on NiFe/IrMn/MgO/Pt tunnel junctions in which large AFM-TAMR of the IrMn/MgO interface is combined with efficient rotation of IrMn staggered moments by exchange-spring effect\cite{Scholl:2004_a} at the opposite interface of IrMn with  the magnetically soft NiFe FM.  

The observed spin-valve-like signal in  our AFM/insulator/normal-metal tunnel junction is shown in Fig.~1a. We detect a stable low resistance state at positive applied magnetic fields, a stable  high resistance state at negative fields, and the switching occurring close to the magnetic field reversal. The measurements where performed on a multilayer structure, sketched in Fig.~1a, deposited by UHV RF sputtering on a  thermally oxidized Si substrate (700nm SiO$_2$ on (001) Si). Multilayers of SiO$_2$/Ta(5)/Ru(10)/Ta(5)/Ni$_{0.8}$Fe$_{0.2}$(10)/Ir$_{0.2}$Mn$_{0.8}$(0, 1.5, 3)/MgO(2.5)/Pt(10) were grown in a magnetic field of 5~mT along the flat edge direction of a 3" wafer (layer thicknesses are given in nm). The structure of the multilayers with (111)-oriented magnetic films was checked by X-ray measurements. Mesa structures of $1\times2\mu$m$^2$ - $5\times10\mu$m$^2$ were patterned from the wafer by photolithography and ion milling. After device fabrication, the wafer was annealed at 350$^{\circ}$C in a magnetic field of 0.4~T applied along the same direction as during the growth. Electrical measurements were done using a four-point contact geometry. Magnetization measurements were performed by the superconducting quantum interference device (SQUID). 

In Fig.~1a we plot measurements on a device fabricated from the wafer with 1.5~nm thick IrMn. We observe a 130\% magnetoresistance effect at 4~K. In Figs.~1b and 1c we compare the magnetoresistance signal ($R(B)$) over a small field range near the reversal with SQUID magnetization ($M(B)$) measurements  on the same wafer with 1.5~nm thick IrMn. The positive/negative field-cooled  magnetization loops provide a clear evidence of the exchange coupling between the NiFe FM and the IrMn AFM. The broadened hysteresis loops of the measured NiFe magnetization, compared to measurements on NiFe without the IrMn AFM in the stack (compare Figs.~1c and 1e), and the opposite horizontal shifts of the  positive and negative field-cooled loops demonstrate the exchange-bias effect\cite{Nogues:1999_a} of IrMn on NiFe. Note that the NiFe FM exchange biased by the IrMn AFM represent a common magnetic electrode used in conventional giant or tunneling magnetoresistance devices. In our structure, however, the order of FM and AFM layers is reversed so that the AFM layer is placed next to the tunnel barrier and governs the transport signal. Similar widths of hysteresis loops of  $R(B)$ and $M(B)$, together with the confirmed NiFe-IrMn exchange coupling, provide the evidence that our structure responds to the applied magnetic field via FM moments in NiFe, that these moments during reversal trigger a tilt of the AFM moments in IrMn (as illustrated in insets of Fig.~1a), and that the tilt of AFM moments results in the strong asymmetric tunneling magnetoresistance signal (Figs~1a and 1b).

Experiments on a control structure without the IrMn layer, presented in Figs.~1d and 1e, highlight the crucial role of the AFM in our study. We emphasize that the control sample has identical structure as shown in the sketch in Fig.~1a, except for the missing IrMn layer, and that the samples with and without IrMn have the same saturation magnetization $M_s=880$~emu/cm$^3$. The hysteretic reversal in the control sample with the pure NiFe magnetic electrode occurs on an order of magnitude smaller scale of applied magnetic fields because NiFe without the exchange biasing AFM is magnetically very soft. We still observe a magnetoresistance signal associated with the switching of magnetization in NiFe, however, it persists only near the small coercive fields and the amplitude  is two orders of magnitude smaller than in the structure with IrMn. The inset of Fig.~1d shows that the amplitude of the low-field magnetoresistance in the field-sweep experiment is the same as the amplitude measured by rotating the sample in a saturating magnetic field. This is a manifestation of the TAMR origin of the measured transport signals in the device with the pure NiFe magnetic electrode.

In Fig.~2 we evidence that in the structure with IrMn placed between NiFe and the MgO tunnel barrier the magnetoresitance effect we observe is also of the TAMR origin. However, since in tunneling devices transport is governed by layers adjacent to the tunnel barrier, we observe  the AFM-TAMR due to IrMn. Both the TAMR and SQUID measurements in Fig.~2 were performed by first applying a field -1~T and then setting the field to a positive value depicted in each panel. By rotating the sample in a 3~mT field we observe zero magnetoresistance (see top left panel in Fig.~2), suggesting that no significant rotation of the NiFe FM moments inside the sample and, therefore, no exchange-spring action on the IrMn AFM is triggered at this low field. Consistently, the projection of the magnetization of the rotated sample to the fixed direction of the applied field measured by SQUID changes from negative to positive values, i.e., the magnetization primarily rotates with the sample and not inside the sample (see top right panel in Fig.~2). At the opposite end of large saturating field of 500~mT (see bottom panels of Fig.~2) we see no change in the magnetization projection with respect to the external field, i.e., the magnetization now rotates inside the sample keeping its orientation parallel to the external field at all angles. The corresponding magnetoresitance shows the low and high resistance states as in the field sweep experiment in Fig.~1a and a smooth transition from one to the other state around 90$^\circ$ angle. This confirms the TAMR origin of the observed magnetoresistance in our AFM/insulator/normal-metal tunnel junction. At intermediate fields there is also a clear correspondence between the SQUID and TAMR data. The magnetization vector depends only weakly on the strength of the applied field near 180$^{\circ}$ field angle and, similarly, the resistance is practically independent of the field strength at these angles. On the other hand, near 0$^{\circ}$ both the magnetization projection and the resistance change with increasing field strength, confirming the TAMR origin of the measured magneto-transport effect. The hysteretic nature of the AFM-TAMR signals reveals that the exchange spring effect of NiFe on IrMn depends on the sense of the rotation of the moments in the structure, i.e. on history, at intermediate fields.

While measurements shown in Fig.~1 and 2 were performed on a NiFe/IrMn/MgO/Pt structure with 1.5~nm thick IrMn, in Fig.~3 we plot magneto-transport and magnetization data obtained on a structure with 3~nm IrMn. The hysteresis loop in the magnetoresistance shown in Fig.~3a is wider for this sample. This is consistent with the enhancement of the AFM-FM exchange-bias  in the 3~nm IrMn sample as observed from the wider and more shifted SQUID magnetization loops. The stronger exchange-bias is also apparent from the larger amplitude of the field which has to be applied to trigger rotation of magnetic moments inside this sample, as shown in Fig.~3b. At 5~mT the 1.5~nm IrMn sample already shows an onset of the TAMR (see Fig.~2) while in the 3~nm IrMn sample no magnetoresistance is detected at this field strength. At 100~mT the moments in NiFe  already tend to align with the external field vector in both samples and, as shown in Fig.~3a and b, the exchange spring effect of NiFe on IrMn remains strong enough also in the 3~nm IrMn structure to yield a large, 160\% spin-valve-like signal due to the AFM-TAMR. 

As an illustration of the rich phenomenology of the FM-AFM exchange coupling effects that can be revealed by the AFM-TAMR we plot in Fig.~4 magnetization and transport data measured at high temperature. From the magnetization data alone, the exchange coupling between IrMn on NiFe seems to have disappeared in our structures at high temperatures. The positive magnetization state at -20~mT (compare Figs.~4a and 1c) is metastable only at low temperatures while at $T\gtrsim 20$~K the magnetization is always aligned with the external field independent of the field sweep direction. Consistently, magnetization hysteresis loops  recorded at temperatures above 20~K show low coercivity and  no horizontal shift, similar to the low-temperature measurement on the sample without IrMn (compare 100~K data in Fig.~4b with 4~K data in Fig.~1e). The AFM-TAMR, however, reveals that the exchange coupling between NiFe and IrMn is still present at high temperatures. As shown in Fig.~4c, we observe a  magnetoresistance at 100~K which is hysteretic on a comparable  range of fields as the high temperature magnetization loop. This is analogous to the 4~K  data shown in Fig.~1b and c, however, the resistances at saturation in the high temperature measurement are similar at positive and negative fields. It suggests that at high temperatures, the exchange-spring effect of NiFe on IrMn yields a more complete rotation of the  staggered moments in IrMn. Figs.~4d-g, showing the TAMR of rotating samples, confirm this observation since a full oscillation period of $R(\theta)$ is detected at high enough fields.  In both the 1.5~nm and 3~nm IrMn samples the TAMR is detected at 500~mT, the former sample showing a 1\% TAMR and the latter sample a 4\% TAMR at 100~K.  At 5~mT, the TAMR is observed only in the 1.5~nm IrMn sample while the moments remain fixed inside the 3~nm IrMn structure. This is consistent with the low temperature measurements which have shown stronger FM-AFM exchange-bias in the 3~nm IrMn sample. 

To conclude, the experimental work presented in this paper builds on our previous theoretical proposal\cite{Shick:2010_a,Jungwirth:2010_a}  of a new approach to spintronics based on relativistic magneto-transport anisotropy phenomena in AFMs. Following the {\em ab initio} predictions of large TAMR in  bi-metallic 3$d$-5$d$ AFM alloys\cite{Shick:2010_a} we prepared tunneling devices with only one magnetic electrode comprising IrMn AFM, separated from a non-magnetic Pt electrode by a MgO barrier. We observe AFM-TAMR signals as large as 160\% at low temperature which is 1-2 orders of magnitude larger than TAMR achieved to date in transition metal FMs and comparable to the tunneling magnetoresistance in conventional spintronic junctions with two FM electrodes. We demonstrate that efficient rotation of staggered moments in the AFM can be induced by the exchange-spring effect of the adjacent FM layer. As a result, our devices show the spin-valve-like magnetoresistance at external fields  $\lesssim50$~mT. With only one magnetic electrode in the tunnel junction we achieved simultaneously large anisotropic magnetoresistance and low switching fields, demonstrating the potential of AFMs for  spintronics. 

Theoretically, there is no apparent physical limit for the spin-orbit induced anisotropic magnetoreistance phenomena to operate at high temperatures. Recall that  the anisotropic magnetoresistance in FM ohmic devices is used in common ambient temperature applications.\cite{Chappert:2007_a} In our pioneering  AFM-TAMR devices, the amplitude of the TAMR is very large at low temperature but at 100~K the effect is reduced to several per cent. Extensive optimization of individual layers and interfaces in the AFM-TAMR structures  is required to establish the limits of high temperature operation. It includes a systematic variation of  parameters like the AFM layer thickness which determine the balance between the robustness of the ordered magnetic state in the AFM and the ability to rotate the staggered moments across the AFM film. The AFM-TAMR itself can be used as a practical experimental tool to investigate these phenomena. If successful, the AFM spintronics concept may represent an attractive alternative to conventional spintronics based on FMs. Particularly appealing is the introduction of AFMs into semiconductor spintronics. Here the synthesis of high Curie temperature FM semiconductors remains a great challenge while AFMs with ordering temperature  safely above room temperature are found\cite{Jungwirth:2010_a} even among magnetic counterparts of the most common semiconductor compounds.

\begin{figure}[h]
\hspace*{-0cm}\epsfig{width=.8\columnwidth,angle=0,file=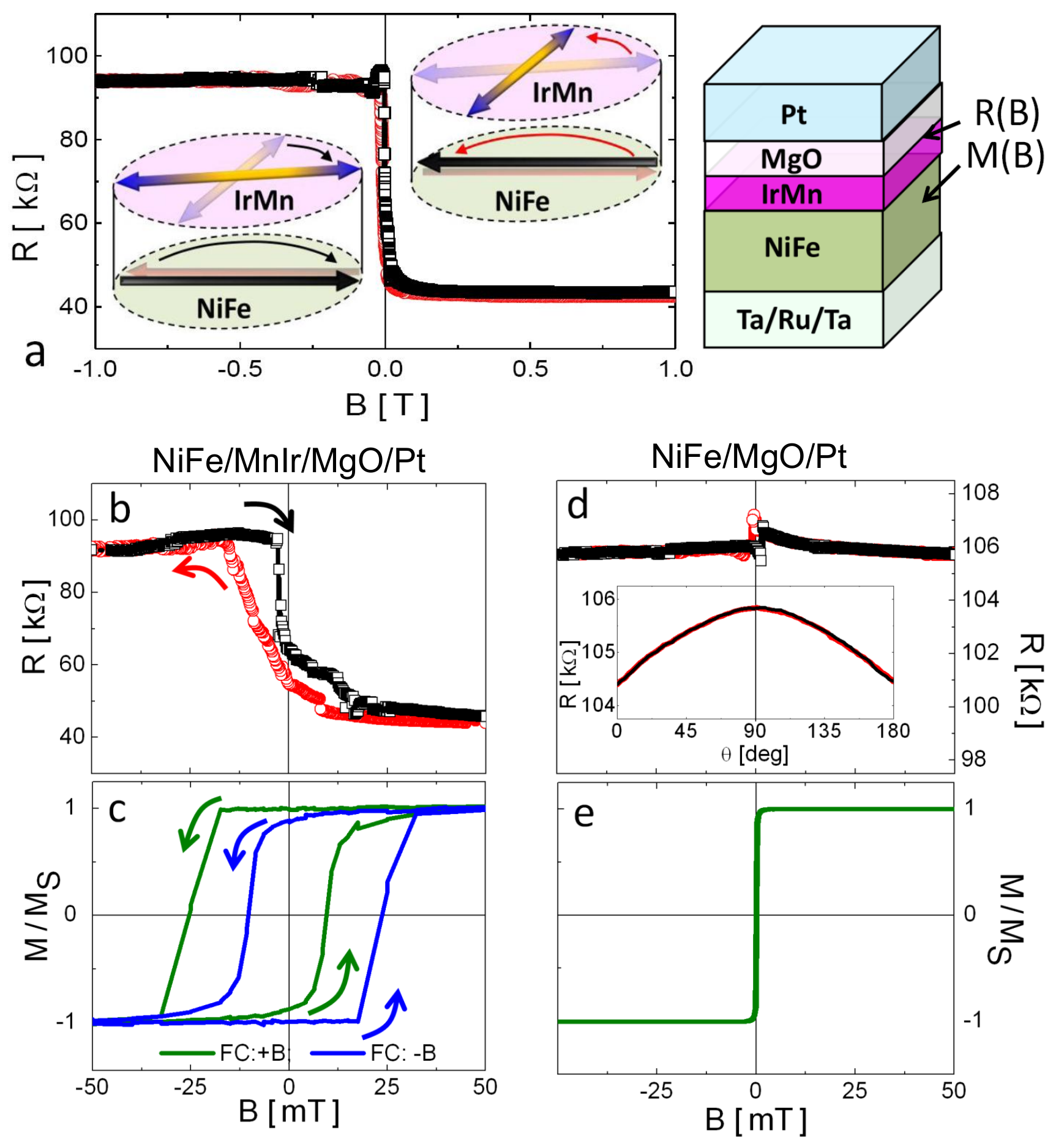}
%
\caption{(a) A spin-valve-like 130\% magnetoresistance signal recorded in the range of -1 to +1~T field on a tunneling device fabricated in the depicted multilayer structure with the NiFe/IrMn(1.5~nm)/MgO/Pt tunnel junction. The insets illustrate the rotation of AFM moments in IrMn via the exchange-spring effect of the adjacent NiFe FM. The external magnetic field is sensed by the NiFe FM while the tunneling transport is governed by the IrMn AFM. (b) Hysteretic magnetoresistance of the NiFe/IrMn(1.5~nm)/MgO/Pt tunnel junction device plotted from -50 to +50~mT. (c) Field-cooled (from 100~K) magnetization loops measured on the same wafer containing the NiFe/IrMn(1.5~nm)/MgO/Pt tunnel junction. (d) Same as (b) measured on a control NiFe/MgO/Pt tunnel junction device; the inset shows the magnetoresistance of the device rotated in a 50~mT field. (e) Magnetization loop of the control wafer without IrMn. All data in the figure were recorded at 4~K.
}
\label{f1}
\end{figure}
\begin{figure}[h]
\hspace*{-0cm}\epsfig{width=.6\columnwidth,angle=0,file=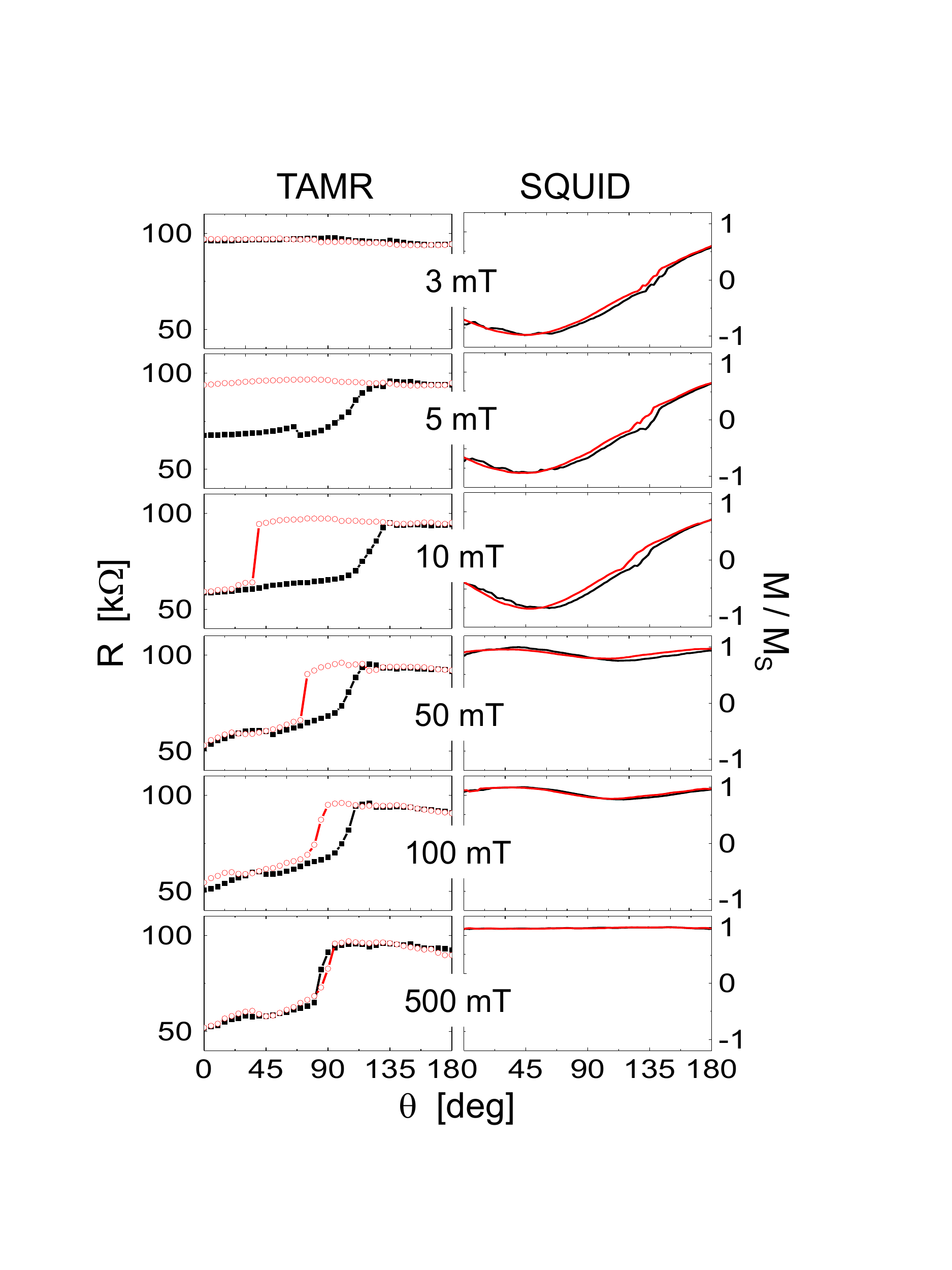}
%

\caption{Left panels: TAMR of the NiFe/IrMn(1.5~nm)/MgO/Pt tunnel junction device rotated in the applied magnetic field of the depicted amplitude between 3 and 500~mT. The sample was rotated first from $\theta=0^{\circ}$ to $180^{\circ}$ (black curves) and then backwards from $180^{\circ}$ to $0^{\circ}$ (red curves). Right panels: corresponding SQUID magnetization measurements of the rotating wafer containing the NiFe/IrMn(1.5~nm)/MgO/Pt tunnel junction. Both the TAMR and SQUID measurements were performed by first applying a field -1~T and then setting the field to the positive value shown in each panel. All data in the figure were recorded at 4~K.
}
\label{f2}
\end{figure}

\begin{figure}[h]

\hspace*{0cm}\epsfig{width=0.9\columnwidth,angle=0,file=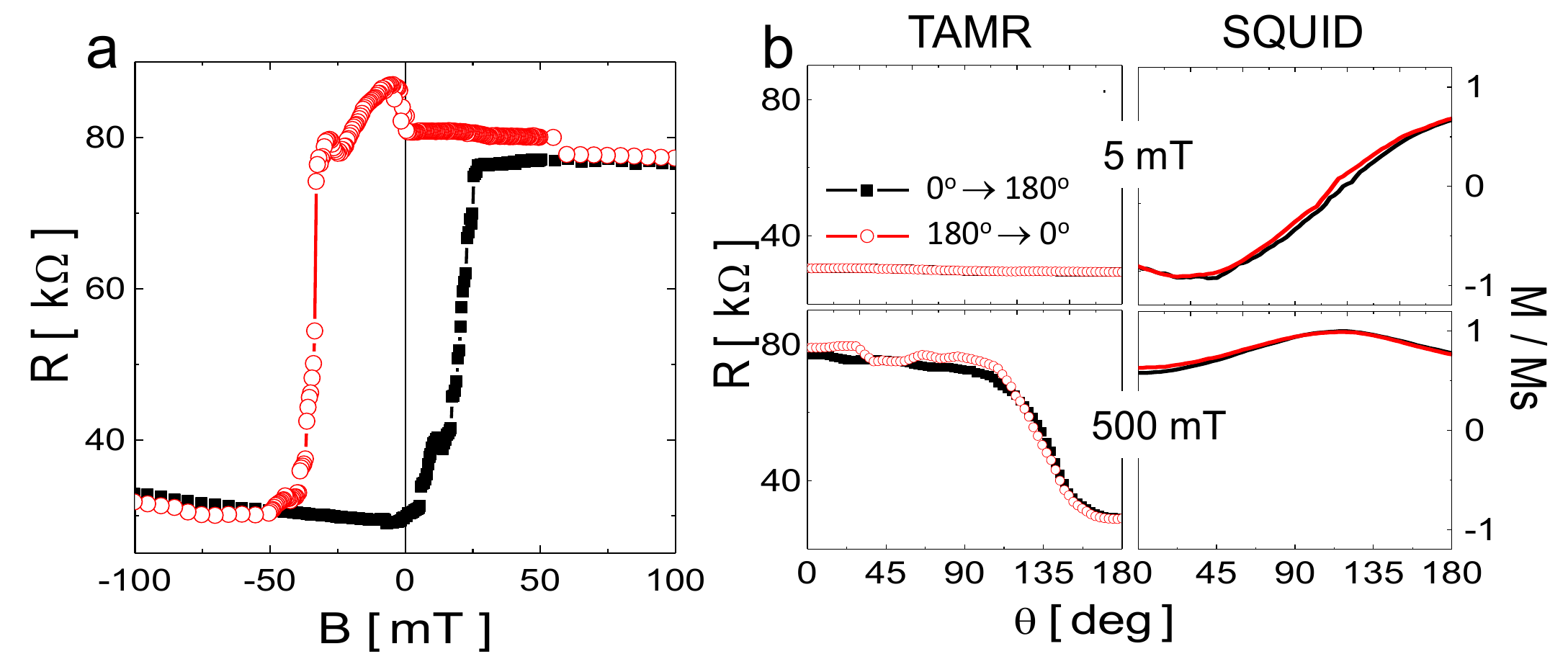}
%

\caption{(a) Hysteretic magnetoresistance of the NiFe/IrMn(3~nm)/MgO/Pt tunnel junction device. The device shows a 160\% spin-valve-like AFM-TAMR. (b) TAMR and SQUID magnetization measurements of the NiFe/IrMn(3~nm)/MgO/Pt tunnel junction device rotated in a 5~mT and 500~mT field. All data in the figure were recorded at 4~K.
}
\label{f2}
\end{figure}

\begin{figure}[h]
\hspace*{-0cm}\epsfig{width=0.9\columnwidth,angle=0,file=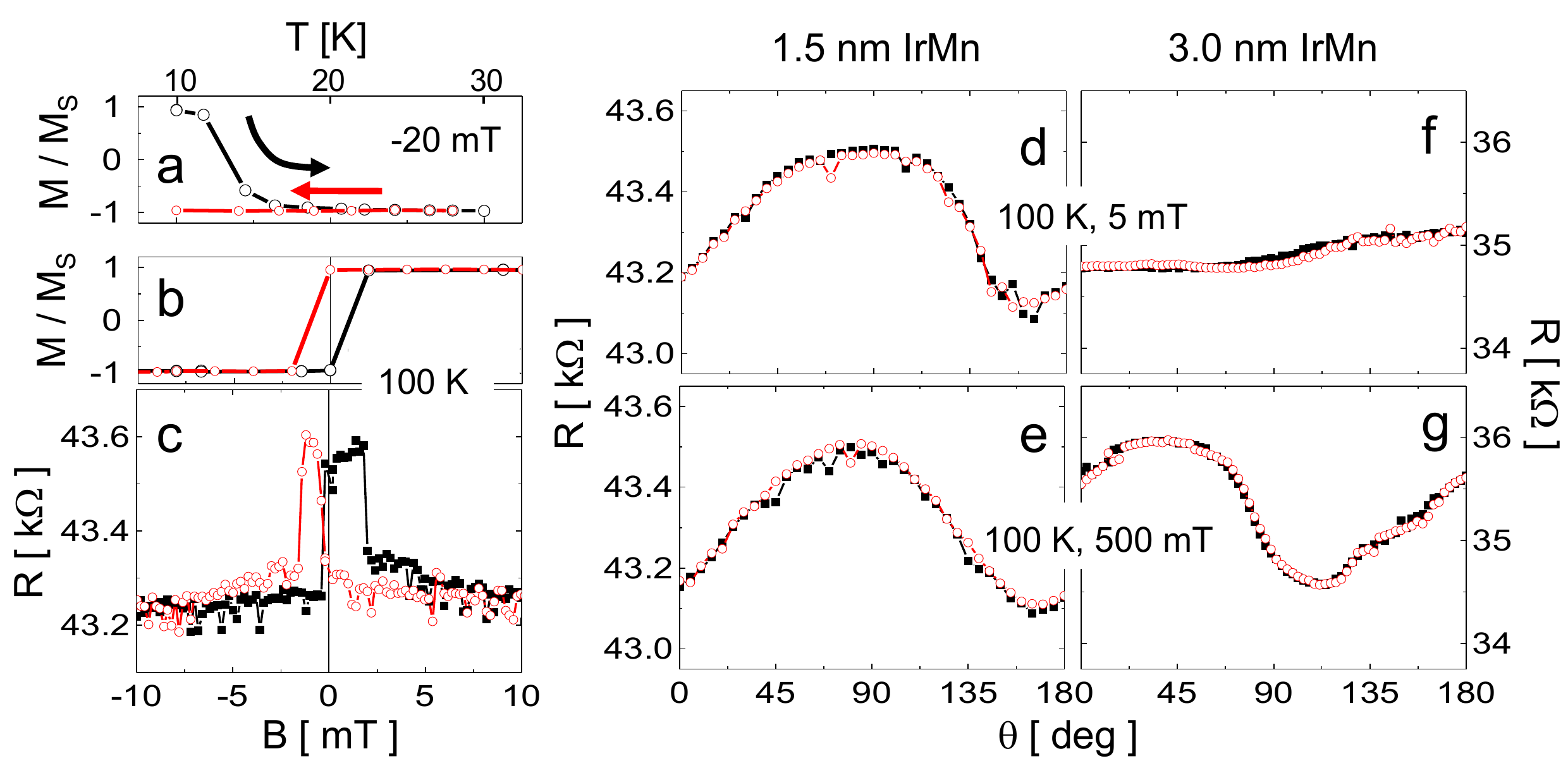}

\vspace*{0cm}
\caption{Temperature dependence of the magnetization of the wafer with 1.5~nm IrMn at -20~mT. At 10~K the magnetization is in the metastable state and it switches to the stable state aligned with the external magnetic field at 17~K. (b) Magnetization loop of the wafer with 1.5~nm IrMn at 100~K. The loop shows no apparent signature of the exchange-bias of NiFe by IrMn. (c) Field-sweep magnetoresistance of the NiFe/IrMn(1.5~nm)/MgO/Pt tunnel junction device at 100~K. (d),(e) TAMR of the rotating NiFe/IrMn(1.5~nm)/MgO/Pt tunnel junction device at 100~K and 5~mT and 500~mT external field, respectively. (f),(g) Same as (d),(e) for the NiFe/IrMn(3~nm)/MgO/Pt tunnel junction device.}
\label{f4}
\end{figure}

\end{document}